\documentclass{article}

 \usepackage[preprint]{neurips_2026}

\usepackage[utf8]{inputenc} 
\usepackage[T1]{fontenc}    
\usepackage{hyperref}       
\usepackage{url}            
\usepackage{booktabs}       
\usepackage{amsfonts}       
\usepackage{nicefrac}       
\usepackage{microtype}      
\usepackage[table]{xcolor}  

\usepackage{graphicx}
\usepackage{multirow}
\usepackage{makecell}
\usepackage{enumitem}
\usepackage{pifont}
\usepackage{subcaption}
\usepackage{rotating}
\newcommand{\cmark}{\ding{51}}
\newcommand{\xmark}{\ding{55}}
\newcommand{\benchname}{\textsc{OxyEcomBench}}

\title{\benchname{}: Benchmarking Multimodal Foundation Models across E-Commerce Ecosystems}

\author{%
  Yong Liu$^{1}$\footnotemark[1], Ximan Liu$^{1}$\footnotemark[1]\hspace{0.5em}\footnotemark[2]\hspace{0.5em}, Guoqing Yang$^{1}$, Bing Bai$^{1}$, Xiaoqiang Xu$^{1}$\\
  \textbf{Zhen Chen$^{1}$, Ke Zhang$^{1}$, Yan Li$^{1}$} \\
  $^{1}$JD.COM \\
  \texttt{liuximan.3@jd.com}
}

\begin{document}

\maketitle
\begingroup
\renewcommand{\thefootnote}{\fnsymbol{footnote}}
\footnotetext[1]{Equal contribution.}
\footnotetext[2]{Project lead.}
\endgroup

\begin{abstract}
Large language models (LLMs) and multimodal large language models (MLLMs) have become indispensable tools across a wide range of applications. E-commerce, however, poses distinctive challenges---including intricate domain knowledge, long-tail product evidence, heterogeneous visual data, and the interplay among multiple stakeholder roles---that diverge substantially from the general world knowledge these models are primarily trained on, often causing a notable gap between their open-domain and e-commerce performance. To systematically quantify this gap, we introduce OxyEcomBench, a unified multimodal benchmark comprising approximately 6,300 high-quality instances for real-world bilingual Chinese--English e-commerce. Although several e-commerce benchmarks have been proposed, they typically adopt a single stakeholder perspective, target a narrow set of tasks, or address isolated challenges (e.g., visual salience or multi-turn dialogues independently), making it difficult to holistically assess models' understanding of the full e-commerce pipeline. OxyEcomBench addresses these limitations by jointly covering platform operators, merchants, and customers across 6 capability aspects and 29 tasks, supporting text-only and mixed-modality inputs with single-image, multi-image, single-turn, and multi-turn configurations. All data is sourced from authentic e-commerce platforms and verified by domain experts. The benchmark further adopts a difficulty-aware design with a four-level P0--P3 rubric applied to all 29 tasks whose difficulty admits stable expert consensus, and rigorously prioritizes visually salient multimodal cases in which key evidence resides in images rather than text alone. Evaluations on 20 mainstream LLMs and MLLMs show that even the leading models attain modest performance and that performance gaps narrow on OxyEcomBench, suggesting that insufficient e-commerce-specific knowledge infusion mutes the advantages of advanced general-purpose models in this domain.

\end{abstract}

\section{Introduction}

Large language models (LLMs)~\citep{achiam2023gpt4} and multimodal large language models (MLLMs)~\citep{liu2023llava,bai2023qwenvl} have demonstrated remarkable capabilities across a wide range of tasks. However, deploying them in highly specialized and complex real-world scenarios, such as e-commerce, presents distinct challenges. As a data-rich and economically significant domain, e-commerce is characterized by platform-specific policies, long-tail product knowledge, and heterogeneous visual evidence. Critically, the e-commerce ecosystem involves three primary stakeholder roles with fundamentally different needs: \emph{platform operators} who manage marketplace infrastructure, policies, and user experience; \emph{merchants} who list products, handle orders, and manage post-sale operations; and \emph{customers} who search, evaluate, and purchase products. This intricate multi-role interplay, combined with the requirement for deep domain expertise, diverges substantially from the general world knowledge these models are primarily trained on, often causing a notable performance gap in real business workflows.

\begin{figure}[t]
    \centering
    \includegraphics[width=0.8\textwidth]{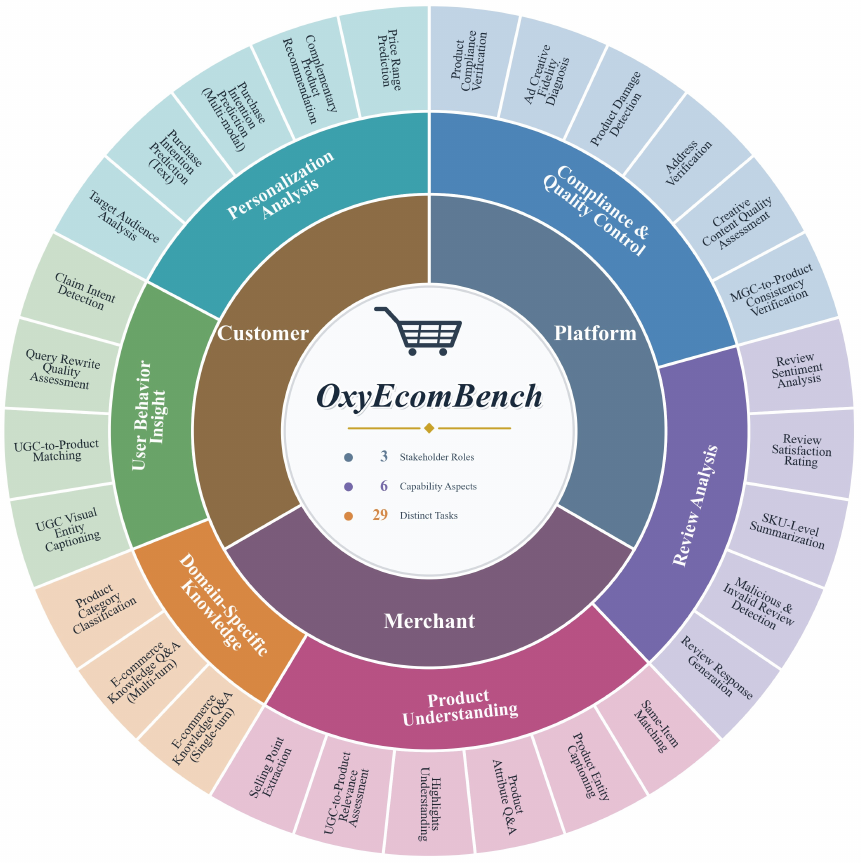}
    \caption{Taxonomy of \benchname{}. The inner ring represents the three stakeholder roles---platform, merchant, and customer---as independent evaluation perspectives. The middle ring groups the benchmark into 6 capability aspects, and the outer ring enumerates 29 distinct e-commerce tasks, with each task inheriting the color of its corresponding capability aspect.}
    \label{fig:overview}
    \vspace{-15pt}
\end{figure}

Recent efforts have made commendable progress in benchmarking LLMs and MLLMs for e-commerce, yet they typically focus on isolated dimensions of the domain. For instance, several benchmarks predominantly adopt a single stakeholder perspective, such as consumer-oriented tasks~\citep{jin2024shoppingmmlu} or customer support scenarios~\citep{chen2025ecombench,sabu2025ecomeval}. Concurrently, other pioneering works have begun to explore specific technical challenges, such as multi-turn conversational interactions~\citep{lu2025rair} or the necessity of visual salience in multimodal tasks. While these targeted evaluations provide valuable insights, they remain largely fragmented. Real-world e-commerce workflows rarely occur in isolation; they frequently involve complex, multi-turn dialogues coupled with visually salient evidence, spanning the diverse and intertwined needs of platform operators, merchants, and customers.

Consequently, the community still lacks a holistic evaluation framework that captures the full complexity of the e-commerce ecosystem. To the best of our knowledge, existing benchmarks largely fall short of simultaneously addressing the distinct evaluation requirements of multiple stakeholder roles within a unified framework. Furthermore, many existing datasets lack fine-grained, expert-calibrated difficulty annotations, making it challenging to determine whether a model's success stems from genuine multi-step reasoning and cross-modal integration, or merely from solving simplistic, surface-level queries.

To systematically quantify the gap between open-domain capabilities and specialized e-commerce requirements, we present \benchname{}. Unlike previous datasets that target isolated challenges, \benchname{} is designed to holistically assess models' understanding of the full e-commerce pipeline. As summarized in Figure~\ref{fig:overview}, the benchmark organizes 3 stakeholder roles, 6 capability aspects, and 29 tasks into a unified taxonomy for real-world e-commerce evaluation. It comprises approximately 6,300 high-quality, bilingual Chinese--English instances spanning these capability aspects and tasks, with each instance manually inspected by e-commerce experts. The dataset rigorously prioritizes visually salient multimodal cases and authentic multi-turn interactions, with all data sourced from real-world e-commerce platforms to ensure ecological validity.

Our main contributions are as follows:
\begin{itemize}[leftmargin=*,nosep]
    \item \textbf{Comprehensive Multi-Role Evaluation:} We introduce \benchname{}, the first benchmark to holistically integrate platform, merchant, and customer perspectives, enabling a complete assessment of foundation models across the entire e-commerce ecosystem.
    \item \textbf{Rigorous Multimodal \& Interactive Assessment:} By curating visually salient tasks and authentic multi-turn dialogues, the benchmark effectively prevents text-only shortcuts and evaluates models on complex, real-world multimodal reasoning.
    In addition, we establish a reliable four-level (P0--P3) difficulty rubric based on expert consensus, providing a nuanced tool to differentiate between surface-level pattern matching and deep cognitive reasoning.
    \item \textbf{Domain-Specific Empirical Insights:} Extensive evaluations on mainstream LLMs and MLLMs show that even leading models attain modest performance and that performance gaps narrow on \benchname{}, indicating that insufficient e-commerce-specific knowledge infusion can mute the advantages of advanced general-purpose models.
\end{itemize}
\section{Related Work}

\subsection{General Multimodal Benchmarks}
The rapid progress of multimodal large language models (MLLMs) has been mirrored by a rich landscape of evaluation suites, ranging from early perception and single-hop reasoning benchmarks (VQA~\citep{antol2015vqa}, COCO Captions~\citep{chen2015cococaptions}, OK-VQA~\citep{marino2019okvqa}, TextVQA~\citep{singh2019textvqa}) and comprehensive multi-task suites (MME~\citep{fu2023mme}, MMBench~\citep{liu2024mmbench}, MM-Vet~\citep{yu2023mmvet}, TouchStone~\citep{bai2023touchstone}, LVLM-eHub~\citep{xu2023lvlm}, LAMM~\citep{yin2023lamm}), to expert-level reasoning tests (ScienceQA~\citep{lu2022scienceqa}, MathVista~\citep{lu2024mathvista}, MMMU~\citep{yue2024mmmu}, CMMU~\citep{he2024cmmu}, ChartQA~\citep{masry2022chartqa}), multi-image and interleaved suites (BLINK~\citep{fu2024blink}, Mantis~\citep{jiang2024mantis}, MUIRBench~\citep{wang2024muirbench}), and reliability-oriented diagnostics targeting hallucination and \emph{visual salience} (POPE~\citep{li2023pope}, HallusionBench~\citep{guan2023hallusionbench}, SEED-Bench~\citep{li2024seedbench}, MMStar~\citep{chen2024mmstar}, MMHal-Bench~\citep{sun2023mmhal}, AMBER~\citep{wang2023amber}). While these general-domain benchmarks offer broad coverage of world knowledge, multi-image reasoning, and visual dependency, they lack the specialized domain expertise---platform-specific policies, long-tail product attributes, and multi-role interactions---required for industrial e-commerce, so strong general-benchmark performance does not necessarily translate into reliability on real-world e-commerce scenarios.

\subsection{E-commerce Specific Benchmarks}
Recognizing the limits of general-domain evaluations, the community has introduced benchmarks tailored to the e-commerce domain. However, these pioneering efforts typically focus on isolated dimensions of the ecosystem, leaving significant gaps in holistic evaluation.

\textbf{Concept Understanding and Text-centric Interactions.} Early e-commerce benchmarks predominantly focused on text-based concept comprehension and single-turn interactions. For instance, Shopping MMLU~\citep{jin2024shoppingmmlu} introduced a comprehensive suite for evaluating consumer-oriented tasks, such as concept understanding and purchase behavior alignment. Similarly, ChineseEcomQA~\citep{liu2025chineseecomqa} focused on foundational e-commerce concept QA. To evaluate interactive capabilities, ECom-Bench~\citep{chen2025ecombench} assessed LLM agents within customer support scenarios, while Mix-Ecom~\citep{zhou2025mixecom} explored mixed-type dialogues and complex domain rules, revealing that current agents struggle with hallucination when navigating intricate e-commerce policies. While valuable, these benchmarks are largely text-centric and overlook the visually rich reality of online shopping.

\textbf{Multimodal Explorations in E-commerce.} Recent works have begun incorporating visual modalities into e-commerce evaluation. EcomEval~\citep{sabu2025ecomeval} integrates a multimodal subset to evaluate basic vision-language alignment. EcomMMMU~\citep{ling2025ecommmu} highlights the necessity of \emph{visually salient} subsets, showing that product images can sometimes introduce redundancy rather than useful signals. Other targeted benchmarks evaluate multimodal representation learning (MBE~\citep{moon2025mbe}), missing modality completion (MMPCBench~\citep{mmpcbench2026}), tool-use in multimodal environments (EcomBench~\citep{min2025ecombench}), visually salient relevance assessment (RAIR~\citep{lu2025rair}), and stage-wise evaluation protocols (EComStage~\citep{zhao2026ecomstage}). Despite these valuable targeted explorations, existing benchmarks remain fragmented, which motivates \benchname{} to provide a unified, full-role, and visually salient evaluation suite.

\begin{table}[t]
\centering
\caption{Comparison of \benchname{} with representative e-commerce benchmarks. \textbf{Roles}: P=Platform operator, M=Merchant, C=Customer. \textbf{Mod.}: T=Text-only, T+V=both text and visual variants. \textbf{M-Turn}: native multi-turn dialogue support. \textbf{M-Img}: multi-image inputs. \textbf{V.Sal.}: deliberate inclusion and emphasis of visually salient tasks to evaluate genuine visual understanding. \textbf{Diff.}: explicit difficulty annotation. \textbf{Reliability} characterizes data trustworthiness: \textbf{Real} indicates task instances are directly sourced from authentic real-world e-commerce platforms (rather than crowd-sourced, reformulated, or LLM-generated); \textbf{Verif.} indicates rigorous human verification by e-commerce domain experts. ``$-$'' denotes that the property is not reported by the original paper.}
\label{tab:benchmark_comparison}
\small
\setlength{\tabcolsep}{3.5pt}
\renewcommand{\arraystretch}{1.1}
\resizebox{\textwidth}{!}{%
\begin{tabular}{lrrccccccccccc c}
\toprule
\multirow{2}{*}{\textbf{Benchmark}} & \multirow{2}{*}{\textbf{\# Tasks}} & \multirow{2}{*}{\textbf{\# Inst.}} & \multicolumn{3}{c}{\textbf{Roles}} & \multicolumn{5}{c}{\textbf{Capabilities}} & \multicolumn{2}{c}{\textbf{Reliability}} & \multirow{2}{*}{\textbf{Lang.}} \\
\cmidrule(lr){4-6} \cmidrule(lr){7-11} \cmidrule(lr){12-13}
 & & & \textbf{P} & \textbf{M} & \textbf{C} & \textbf{Mod.} & \textbf{M-Turn} & \textbf{M-Img} & \textbf{V.Sal.} & \textbf{Diff.} & \textbf{Real} & \textbf{Verif.} & \\
\midrule
WebShop~\citep{yao2022webshop}            & 1   & 12K    & \xmark & \xmark & \cmark & T+V & \cmark & \xmark & \xmark & \xmark & \cmark & \xmark & EN \\
Shopping MMLU~\citep{jin2024shoppingmmlu} & 57  & 20.8K  & \xmark & \xmark & \cmark & T   & \xmark & \xmark & \xmark & \xmark & \cmark & \xmark & Multi \\
ChineseEcomQA~\citep{liu2025chineseecomqa}& 10  & 1.8K   & \xmark & \xmark & \cmark & T   & \xmark & \xmark & \xmark & \xmark & \xmark & \cmark & ZH \\
ECom-Bench~\citep{chen2025ecombench}      & 53  & -     & \xmark & \xmark & \cmark & T+V & \cmark & \xmark & \xmark & \xmark & \cmark & \cmark & ZH \\
EcomMMMU~\citep{ling2025ecommmu}          & 8   & 406K   & \xmark & \xmark & \cmark & T+V & \xmark & \cmark & \cmark & \xmark & \cmark & \xmark & EN \\
EcomEval~\citep{sabu2025ecomeval}         & 37  & 3.1K   & \xmark & \cmark & \cmark & T+V & \xmark & \xmark & \xmark & \cmark & \cmark & \cmark & 7 langs \\
Mix-ECom~\citep{zhou2025mixecom}          & 4   & 4.8K   & \xmark & \xmark & \cmark & T+V & \cmark & \cmark & \xmark & \xmark & \cmark & \cmark & ZH \\
EcomBench~\citep{min2025ecombench}        & 7   & $-$    & \xmark & \cmark & \cmark & T+V & \cmark & \xmark & \xmark & \cmark & \cmark & \cmark & EN \\
RAIR~\citep{lu2025rair}                   & 14   & 49k    & \xmark & \xmark & \cmark & T+V & \xmark & \xmark & \cmark & \cmark & \cmark & \cmark & ZH \\
EComStage~\citep{zhao2026ecomstage}       & 7   & 4.8K   & \xmark & \cmark & \cmark & T+V & \cmark & \xmark & \xmark & \xmark & \cmark & \cmark & ZH+EN \\
\midrule
\rowcolor{gray!15}
\textbf{\benchname{} (Ours)}              & \textbf{28} & \textbf{6.3K} & \cmark & \cmark & \cmark & \textbf{T+V} & \cmark & \cmark & \cmark & \cmark & \cmark & \cmark & \textbf{ZH+EN} \\
\bottomrule
\end{tabular}%
}
\end{table}

Despite these commendable advancements, the existing landscape remains fragmented. As summarized in Table~\ref{tab:benchmark_comparison}, most benchmarks adopt a single stakeholder perspective (e.g., consumer-oriented or customer-support) and fail to integrate the distinct requirements of platform operators, merchants, and customers; few simultaneously enforce visual salience, multi-turn interactions, and expert-calibrated difficulty within a bilingual setting. \benchname{} bridges this critical gap. By combining multi-role coverage, emphasis on visually salient tasks (preventing text-only shortcuts from dominating), authentic multi-turn interactions, and expert-calibrated difficulty levels within a bilingual framework, \benchname{} provides the first holistic testbed for evaluating foundation models across the entire e-commerce ecosystem.
\section{\benchname{}}

\benchname{} is a unified bilingual multimodal benchmark for evaluating foundation models in real-world e-commerce. It comprises approximately 6{,}300 high-quality Chinese--English instances across 29 tasks, 6 capability aspects, and 3 stakeholder roles---platform operators, merchants, and customers---covering text-only, single-image, multi-image, single-turn, and multi-turn settings. All instances are sourced from authentic e-commerce platforms and assigned deterministic task-level P0--P3 difficulty labels. Their reference answers, derived from a hybrid of authentic human feedback and high-quality model generations, undergo systematic verification by powerful models and senior domain experts to ensure accuracy, reliability, and ecological validity.

\subsection{Design Principles}
\label{sec:design}

The design of \benchname{} is guided by three core principles:

\textbf{Full-role coverage.} E-commerce ecosystems involve three distinct stakeholder roles---\emph{platform operators} (governing marketplace policy and content quality), \emph{merchants} (handling product listing, quality inspection, and operational decisions), and \emph{customers} (searching, consulting, and purchasing). \benchname{} jointly covers all three roles within a unified evaluation suite, enabling holistic assessment across the full e-commerce workflow rather than optimization for a single perspective as some previous works.

\begin{figure}[t]
    \centering
    \includegraphics[width=\textwidth]{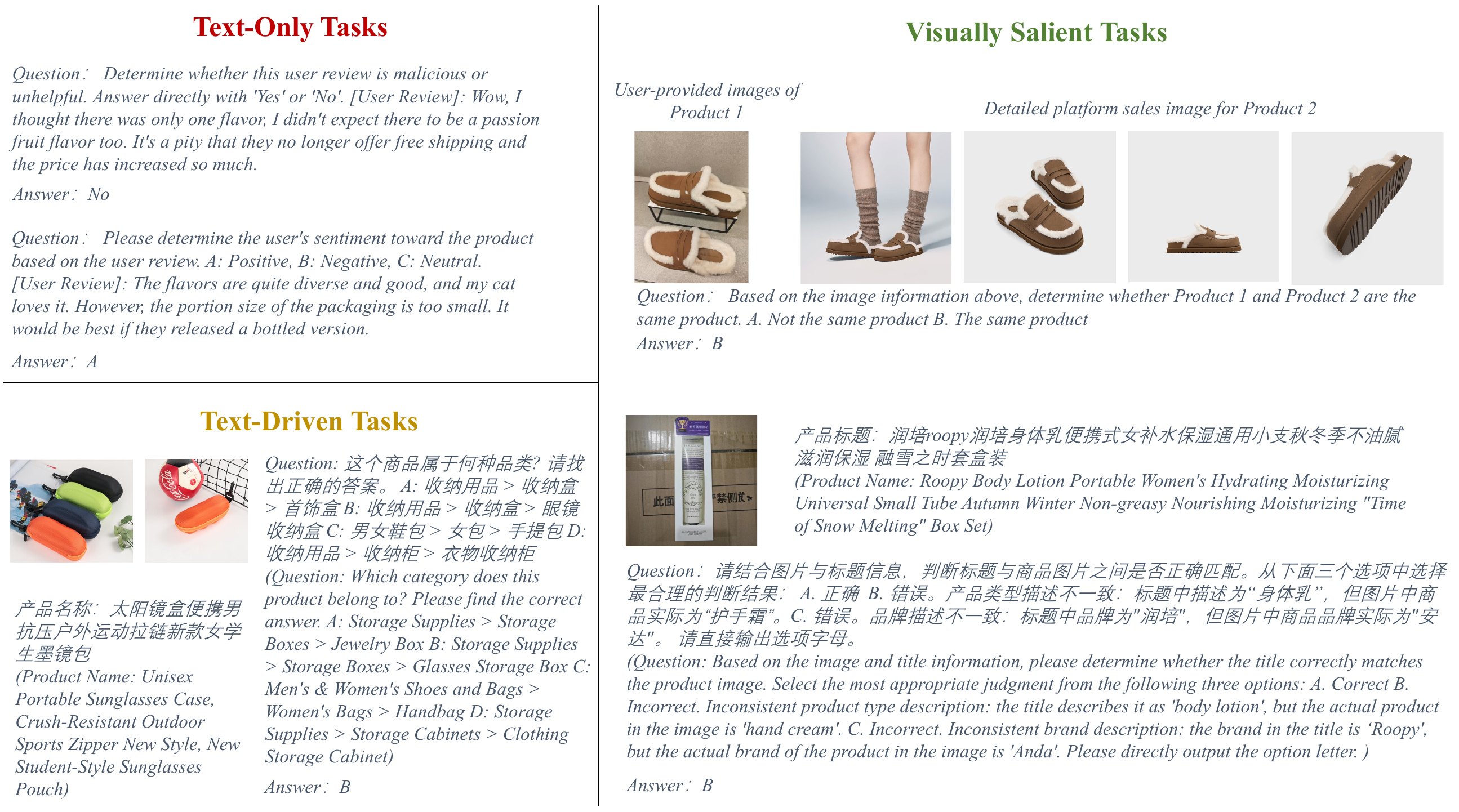}
    \caption{Illustration of three types of tasks in \benchname{}. \emph{Text-Only Tasks:} instances that rely purely on textual information, such as review classification or sentiment analysis. \emph{Text-Driven Tasks:} multimodal tasks where the answer can be derived primarily from textual cues. \emph{Visually Salient Tasks:} cases where fine-grained visual evidence (e.g., matching product details across different environments) is indispensable for the correct answer.}
    \label{fig:visual_salience}
\end{figure}

\textbf{Visual salience.} Images serve as one of the most critical information carriers in e-commerce scenarios. To rigorously evaluate models' understanding of visual content, we treat a multimodal task as \emph{visually salient} when correct responses depend on visual attributes (appearance, fine-grained defects, or cross-image correspondence) rather than text alone. By curating a rich collection of visually salient instances, \benchname{} ensures that models must genuinely process and integrate visual evidence to succeed, preventing text-only shortcuts from dominating and inflating multimodal benchmark scores~\citep{ling2025ecommmu,lu2025rair}. Figure~\ref{fig:visual_salience} illustrates our three task modalities, highlighting our focus on visually salient tasks where fine-grained visual evidence is indispensable for correct reasoning.

\textbf{Difficulty-aware categorization.} A four-level rubric (P0--P3, from lowest to highest difficulty) classifies tasks across three dimensions: cognitive complexity, required information integration, and domain knowledge depth. P0 (\emph{Fundamental Understanding \& Explicit Extraction}) requires basic single-modality comprehension and explicit information extraction; P1 (\emph{Intermediate Analysis \& Routine Reasoning}) demands context integration and routine domain reasoning; P2 (\emph{Advanced Reasoning \& Fine-grained Comparison}) necessitates fine-grained cross-modal comparison, multi-image reasoning, implicit intention mining, and sustained context tracking in multi-turn dialogues; and P3 (\emph{Expert-Level Judgment \& Comprehensive Decision-Making}) involves deep domain knowledge, complex policy application, and comprehensive multi-step decisions. To maintain consistency and rigorous evaluation, all 29 tasks receive deterministic task-level difficulty labels, which were determined by multiple e-commerce domain experts through comprehensive analysis of actual instances and consensus voting. This rigorous process preserves cross-task comparability and label reliability. The specific difficulty level assigned to each task is detailed in Table~\ref{tab:taxonomy}.

\begin{table}[t]
    \centering
    \caption{Task taxonomy of \benchname{}. The Role column lists the three stakeholder perspectives without horizontal subdivision, while horizontal rules start from the Aspect column to separate capability aspects. \textbf{Mod.}: T = Text-only, T+I = Text + Single Image, T+MI = Text + Multiple Images. The multimodal (T+I or T+MI) tasks emphasize visually salient data to evaluate genuine visual understanding.}
    \label{tab:taxonomy}
    \small
    \resizebox{\textwidth}{!}{%
    \begin{tabular}{p{1.4cm}p{3cm}p{5.8cm}ccc}
    \toprule
    \textbf{Role} & \textbf{Aspect} & \textbf{Tasks} & \textbf{\# Samples} & \textbf{Mod.} & \textbf{Difficulty} \\
    \midrule
     & \multirow{6}{*}{\makecell[l]{Compliance \&\\Quality Control}} & Product Compliance Verification & 199 & T+I & P2 \\
     & & Ad Creative Fidelity Diagnosis & 160 & T+MI & P2 \\
     & & Product Damage Detection & 177 & T+I & P1 \\
    \textbf{Platform} & & Address Verification & 100 & T & P1 \\
     & & Creative Content Quality Assessment & 160 & T+MI & P2 \\
     & & MGC-to-Product Consistency Verification & 200 & T+MI & P3 \\
    \cmidrule{2-6}
     & \multirow{5}{*}{Review Analysis} & Review Sentiment Analysis & 429 & T & P1 \\
     & & Review Satisfaction Rating & 430 & T & P1 \\
     & & SKU-Level Summarization & 200 & T & P1 \\
     & & Malicious \& Invalid Review Detection & 245 & T & P0 \\
     & & Review Response Generation & 59 & T+MI & P2 \\
    \cmidrule{2-6}
     & \multirow{6}{*}{Product Understanding} & Same-Item Matching & 238 & T+MI & P2 \\
     & & Product Entity Captioning & 220 & T+MI & P2 \\
     & & Product Attribute Q\&A & 150 & T+MI & P2 \\
     & & Highlights Understanding & 168 & T+I & P2 \\
    \textbf{Merchant} & & UGC-to-Product Relevance Assessment & 296 & T+MI & P1 \\
     & & Selling Point Extraction & 214 & T+MI & P1 \\
    \cmidrule{2-6}
     & \multirow{3}{*}{\makecell[l]{Domain-Specific\\Knowledge}} & E-commerce Knowledge Q\&A (Single-turn) & 589 & T & P0 \\
     & & E-commerce Knowledge Q\&A (Multi-turn) & 233 & T & P2 \\
     & & Product Category Classification & 112 & T+MI & P1 \\
    \cmidrule{2-6}
     & \multirow{4}{*}{\makecell[l]{User Behavior\\Insight}} & UGC Visual Entity Captioning & 212 & T+I & P2 \\
     & & UGC-to-Product Matching & 287 & T+MI & P2 \\
     & & Query Rewrite Quality Assessment & 173 & T & P1 \\
     & & Claim Intent Detection & 200 & T & P1 \\
    \cmidrule{2-6}
    \textbf{Customer} & \multirow{5}{*}{\makecell[l]{Personalization\\Analysis}} & Target Audience Analysis & 200 & T+MI & P1 \\
     & & Purchase Intention Prediction (Text) & 176 & T & P3 \\
     & & Purchase Intention Prediction (Multi-modal) & 139 & T+MI & P3 \\
     & & Complementary Product Recommendation & 155 & T+MI & P1 \\
     & & Price Range Prediction & 193 & T+MI & P2 \\
    \bottomrule
    \end{tabular}%
    }
\end{table}

\subsection{Task Taxonomy}
\label{sec:taxonomy}

\benchname{} organizes 29 tasks across 6 capability aspects, reflecting the intertwined needs of the three stakeholder roles (Table~\ref{tab:taxonomy}). In task names, UGC and MGC denote user- and merchant-generated content, respectively. The tasks test core abilities: \emph{Compliance \& Quality Control} examines rule following, inspection, and image-text consistency; \emph{Review Analysis} evaluates opinion understanding, summarization, and response generation; \emph{Product Understanding} covers product identity, attributes, relevance, and selling-point extraction; \emph{Domain-Specific Knowledge} tests e-commerce Q\&A and product classification; \emph{User Behavior Insight} focuses on real-world recognition, matching, query rewriting, and intent detection; and \emph{Personalization Analysis} evaluates audience analysis, purchase prediction, recommendation, and pricing assessment.

\begin{figure}[t]
    \centering
    \includegraphics[width=\textwidth]{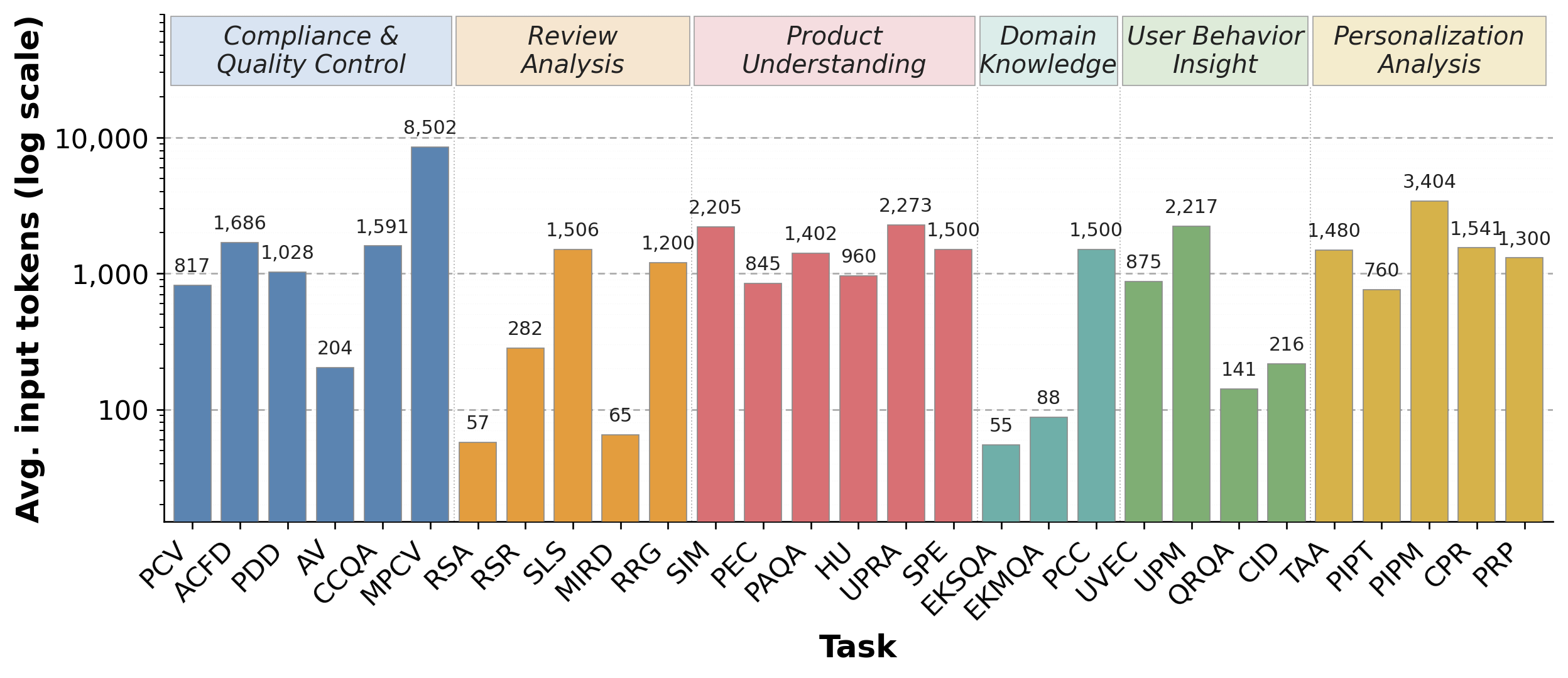}
    \vspace{-15pt}
    \caption{Per-task average input-token length of \benchname{} (log scale). Bars are grouped by capability aspect (top band); task names use the abbreviations defined in Table~\ref{tab:taxonomy}. Average inputs span nearly two orders of magnitude---from $\approx$50 tokens for short query-style tasks (e.g., \emph{QRQA}) to $\approx$8{,}500 tokens for reasoning- and dialogue-intensive tasks (e.g., \emph{EKMQA})---enabling joint evaluation of short-, mid-, and e-commerce-specific long-context capabilities.}
    \label{fig:token_length_distribution}
\end{figure}

To comprehensively probe these capabilities, \benchname{} accommodates diverse task formats. \emph{Question types} include single-choice, multiple-choice, true/false, and open-ended Q\&A. \emph{Dialogue formats} include single-turn tasks and multi-turn tasks that simulate conversational interactions (e.g., customer service), evaluating the model's ability to maintain context across exchanges. \emph{Modality configurations} span text-only, single-image, and multi-image inputs. Our multimodal tasks emphasize visual salience as described in Section~\ref{sec:design}.
Furthermore, as illustrated in Figure~\ref{fig:token_length_distribution}, the per-task average input-token length spans nearly two orders of magnitude---from short, query-style inputs of around 50 tokens to context-heavy inputs that approach 8{,}500 tokens. While modern foundation models can process significantly longer sequences in general domains (e.g., summarizing 100K-token books), 8{,}500 tokens in e-commerce represents a highly information-dense, long-context scenario. It typically encompasses extensive multi-turn customer service histories with complex coreference, or intricate platform policy documents requiring strict logical adherence. Such breadth lets \benchname{} jointly probe short-context understanding, mid-range comprehension, and e-commerce-specific long-context reasoning within a single benchmark.

\subsection{Data Construction}
\label{sec:construction}

To guarantee high data quality and ecological validity, the construction of \benchname{} follows a systematic three-stage pipeline designed to source authentic data, challenge models via hard negative mining, and ensure reliability.

\textbf{Stage 1: Real-World Data Collection.} We collect raw data from diverse business scenarios within a large-scale e-commerce platform, including advertising, transactions, search, recommendation, and compliance. The corpus encompasses heterogeneous modalities: product listings with single/multiple images, customer reviews, multi-turn dialogue logs, and complex policy documents. All data undergoes strict anonymization to remove PII and sensitive business metrics.

\textbf{Stage 2: Data Processing \& Annotation.} After filtering noisy or incomplete records, we design task-specific questions and interaction formats. For instances with natural ground truth, we construct \emph{hard negative samples} to mimic realistic pitfalls. For example, in product understanding, we subtly alter textual attributes to misalign with the product; in matching tasks, we substitute images with highly similar but distinct products. For instances lacking explicit ground truth, we employ a collaborative paradigm where initial model-generated annotations are refined by human annotators following detailed guidelines.

\textbf{Stage 3: Quality Assurance.} We implement a dual-verification mechanism. First, state-of-the-art foundation models (e.g., GPT-5.4~\citep{openai2026gpt5}, Gemini 3.1 Pro~\citep{google2026gemini3}) automatically inspect the rationality of questions and answers against predefined rules. Finally, senior e-commerce domain experts conduct a comprehensive manual review to verify factual correctness, policy alignment, and absence of ambiguity, thereby guaranteeing the ultimate accuracy of the evaluation instances.

\section{Experiments}
\label{sec:experiments}


\subsection{Experimental Setup}
\label{sec:setup}

\textbf{Evaluated models.}
We benchmark $20$ representative LLMs and MLLMs that span a diverse landscape of architectures, scales, and openness.
\textbf{Proprietary models} include \emph{Gemini 3 Pro} and \emph{Gemini 3 Flash}, \emph{GPT-5.2} and \emph{GPT-4o}, and \emph{Seed 2.0 Pro}.
\textbf{Open-source / open-weight models} include \emph{Kimi 2.5}; \emph{Qwen3.5-397B-A17B}, \emph{Qwen3.5-35B}, \emph{Qwen3.5-27B}, \emph{Qwen3.5-9B}, and \emph{Qwen3.5-2B}; \emph{Qwen3-VL-32B} and \emph{Qwen3-VL-8B}; \emph{InternVL3.5-8B}; \emph{GLM-5.1} and \emph{GLM-5.0}; \emph{DeepSeek-V4-Flash}; \emph{Llama3.3-70B}, \emph{Llama3.1-8B}, and \emph{Llama3.2-11B-Vision}.
This suite jointly covers \emph{multimodal} models (Gemini 3 series, GPT-4o/5.2, Seed 2.0 Pro, Kimi 2.5, Qwen3.5 series, Qwen3-VL series, InternVL3.5, and Llama3.2-Vision) and \emph{text-only} models (GLM-5.1, GLM-5.0, Llama3.3, Llama3.1, and DeepSeek-V4-Flash), enabling controlled analysis of multimodal capability gaps.
For text-only models on multimodal tasks, cells are left blank when no caption-fallback result is provided in the source spreadsheet; aggregate scores are computed over the available task results.

\textbf{Evaluation metrics.}
We adopt task-specific metrics following standard practices. Objective formats (true/false, single-/multiple-choice) are evaluated using \emph{Accuracy}. For open-ended Q\&A, we employ an LLM-as-judge protocol, utilizing a strong third-party model to assess responses based on a carefully designed grading rubric. All per-task scores are normalized to $[0,100]$. In Table~\ref{tab:main_results}, per-capability scores and the \emph{Overall} score are macro averages over the available task scores within each aspect and across all tasks, respectively.
All models are evaluated zero-shot with manually designed task-specific prompts. Open-weight models are run locally or via compatible hosted endpoints; proprietary models are accessed via official APIs.

\subsection{Main Results}
\label{sec:main_results}

\begin{table}[t]
    \centering
    \caption{Main results on \benchname{} across the six capability aspects. \emph{C\&QC}: Compliance \& Quality Control; \emph{RA}: Review Analysis; \emph{PU}: Product Understanding; \emph{DK}: Domain-Specific Knowledge; \emph{UBI}: User Behavior Insight; \emph{PA}: Personalization Analysis. Per-capability and overall scores are macro averages over the corresponding non-missing task scores in Table~\ref{tab:per_task_results}. Scores for text-only Large Language Models (marked with $^*$) are computed only on text-only tasks and are therefore not directly comparable to Vision-Language Models with full multimodal task coverage. \textbf{Bold} denotes the best score and \underline{underlined} denotes the second-best score within each column.}
    \label{tab:main_results}
    \footnotesize
    \renewcommand{\arraystretch}{0.9}
    \begin{tabular*}{\textwidth}{@{\extracolsep{\fill}}lcccccc|c@{}}
    \toprule
    \textbf{Model} & \textbf{C\&QC} & \textbf{RA} & \textbf{PU} & \textbf{DK} & \textbf{UBI} & \textbf{PA} & \textbf{Overall} \\
    \midrule
    \rowcolor{gray!15}\multicolumn{8}{c}{\textit{Proprietary Models}} \\
    \midrule
    \multicolumn{8}{l}{\textit{\quad Vision-Language Models (VLM)}} \\
    \cmidrule(lr){1-8}
    Gemini~3~Pro             & 74.5 & \textbf{79.4} & 72.7 & 69.7 & 67.8 & 48.6 & \textbf{69.1} \\
    Gemini~3~Flash           & 74.3 & 76.2 & 72.9 & 69.5 & 71.9 & 47.8 & \underline{68.9} \\
    GPT-5.2                  & 66.6 & 70.3 & 68.0 & \underline{73.7} & 69.3 & 47.3 & 65.3 \\
    GPT-4o                   & 56.9 & 69.4 & 66.0 & 67.4 & 68.7 & 47.3 & 62.0 \\
    Seed~2.0~Pro             & 64.8 & 71.0 & 72.8 & 68.9 & 71.7 & 45.9 & 65.6 \\
    \midrule
    \rowcolor{gray!15}\multicolumn{8}{c}{\textit{Open-source / Open-weight Models}} \\
    \midrule
    \multicolumn{8}{l}{\textit{\quad Vision-Language Models (VLM)}} \\
    \cmidrule(lr){1-8}
    Kimi~2.5                 & 70.6 & 68.9 & \textbf{74.8} & 69.1 & \underline{74.4} & \textbf{49.5} & 67.9 \\
    Qwen3.5-397B-A17B        & 67.7 & 73.5 & \underline{73.7} & 71.5 & 74.0 & \underline{49.5} & 68.0 \\
    Qwen3.5-35B              & 59.3 & 66.1 & 67.2 & 69.6 & 68.6 & 47.4 & 62.4 \\
    Qwen3.5-27B              & 62.4 & 70.6 & 68.8 & 70.9 & 70.5 & 47.2 & 64.5 \\
    Qwen3.5-9B               & 60.6 & 66.7 & 64.9 & 69.4 & 64.4 & 46.5 & 61.5 \\
    Qwen3.5-2B               & 41.2 & 57.9 & 58.3 & 66.4 & 61.5 & 40.2 & 52.9 \\
    Qwen3-VL-32B             & 68.8 & 69.5 & 69.1 & \textbf{85.7} & \textbf{75.2} & 45.7 & 67.6 \\
    Qwen3-VL-8B              & 62.3 & 68.7 & 64.9 & 70.5 & 70.1 & 44.9 & 62.9 \\
    InternVL3.5-8B           & 47.2 & 60.0 & 55.5 & 66.9 & 60.5 & 43.8 & 54.4 \\
    Llama3.2-11B-V           & 45.6 & 56.7 & 48.8 & 63.0 & 45.5 & 38.8 & 48.8 \\
    \cmidrule(lr){1-8}
    \multicolumn{8}{l}{\textit{\quad Large Language Models (LLM)}} \\
    \cmidrule(lr){1-8}
    GLM-5.1$^*$                  & 92.0 & 73.5 & -- & 59.9 & 63.5 & 1.1 & 63.4 \\
    DeepSeek-V4-Flash$^*$        & \underline{92.3} & 67.6 & -- & 60.3 & 57.6 & 3.8 & 60.2 \\
    GLM-5.0$^*$                  & \textbf{99.0} & 74.0 & -- & 58.1 & 61.0 & 1.7 & 63.5 \\
    Llama3.3-70B$^*$             & 88.0 & \underline{77.2} & -- & 56.1 & 62.8 & 5.1 & 64.0 \\
    Llama3.1-8B$^*$              & 58.0 & 61.3 & -- & 55.6 & 58.4 & 2.8 & 53.4 \\
    \bottomrule
    \end{tabular*}
\end{table}

Due to space constraints, Table~\ref{tab:main_results} reports only the per-capability and overall scores of the $20$ evaluated models; detailed per-task performance metrics are provided in Appendix~\ref{appendix:per_task_results}.

\textbf{A low ceiling despite strong general-purpose capabilities.} The highest overall score is only $69.1$ (Gemini~3~Pro), closely followed by Gemini~3~Flash ($68.9$) and the leading open-weight models, Qwen3.5-397B-A17B ($68.0$) and Kimi~2.5 ($67.9$). Unlike saturated leaderboards where a single frontier model dominates, models with vastly different architectures, scales, and open-domain capabilities are compressed into a narrow performance band on real e-commerce workflows. This suggests that general multimodal competence does not transfer cleanly to domain-specific knowledge, long-tail product evidence, and multi-role operational judgments. This validates the core motivation for \benchname{}: broad evaluations obscure capability gaps that only emerge under realistic e-commerce constraints.

\textbf{Leaderboard compression highlights the need for domain-specific knowledge injection.} Open-weight systems rival proprietary models, with Qwen3.5-397B-A17B and Kimi~2.5 ranking third and fourth overall. While the Qwen3.5 family exhibits expected scaling benefits, these gains are modest compared to the remaining gap to reliable e-commerce performance. Thus, \benchname{} diverges from the familiar ordering of general-purpose benchmarks. It highlights that in vertical domains, the marginal gains from purely scaling up parameters diminish. Generic instruction following is insufficient without e-commerce-specific knowledge, visual grounding, and multi-role reasoning, underscoring the necessity of domain-specific post-training. (Note that text-only LLM scores reflect a limited language-only subset rather than full multimodal performance.)

\textbf{No model exhibits a complete e-commerce capability profile.} Among VLMs, no single model dominates all six capability aspects: Gemini~3~Pro leads Review Analysis; Kimi~2.5 leads Product Understanding and Personalization Analysis; and Qwen3-VL-32B leads Domain-Specific Knowledge and User Behavior Insight. These crossovers indicate that e-commerce competence is not a single latent ability. By disentangling distinct capabilities, the benchmark reveals diverse failure modes for otherwise strong models (e.g., misjudging exaggerated merchant claims as factual, or missing fine-grained visual discrepancies across images). While text-only LLMs (e.g., GLM-5.0) score exceptionally well on the text-only subset of Compliance \& Quality Control, they lack the visual capabilities required for the full multimodal quality control workflow, rendering their aspect-level scores incomparable to VLMs.

\subsection{Analysis}
\label{sec:analysis}

\textbf{The hardest failures align with real business decisions.}
Task-level averages span a wide range, from $3.1$ on Purchase Intention Prediction (Text) to $99.0$ on Highlights Understanding. This spread confirms that low-scoring tasks are not artificial corner cases but authentic business bottlenecks. For instance, purchase-intention prediction requires synthesizing noisy user profiles and long-term behaviors. Averaging task means by stakeholder role yields $67.5$ for Merchant-side, $64.3$ for Platform-side, and $55.4$ for Customer-side tasks. Qualitative analysis reveals that while models readily adopt a "merchant persona" to extract explicit attributes, they struggle to infer latent "customer intent," frequently hallucinating preferences unsupported by the context.

\textbf{Per-task leadership reveals fragmented strengths.}
Rankings are distributed across model families rather than dominated by one system: Qwen3.5-397B-A17B ranks in the top two on $11$ tasks, while Gemini~3~Pro and Kimi~2.5 do so on $9$ tasks each. This fragmentation stems from divergent training paradigms: open-weight models often excel in Chinese e-commerce contexts and product understanding, whereas proprietary models (e.g., Gemini) demonstrate stronger complex reasoning. High-scoring tasks (e.g., Highlights Understanding at $99.0$, Product Category Classification at $89.3$) typically involve explicit extraction or isolated knowledge lookup. Conversely, operational tasks like Review Response Generation ($49.7$) demand nuanced response planning and tone control that models currently lack.

\textbf{Fine-grained visual recognition and complex reasoning expose critical failure modes.}
\benchname{} rigorously tests capabilities beyond general priors. In visually salient tasks like MGC-to-Product Consistency Verification (averaging $20.8$), error analysis indicates that models frequently fail to align subtle visual discrepancies---such as minor color variations or missing accessories---with product descriptions, often defaulting to a "consistent" guess. Similarly, in complex reasoning scenarios like E-commerce Knowledge Q\&A (Multi-turn) (averaging $23.3$), models exhibit severe context confusion. Rather than accurately synthesizing information across multiple dialogue turns to address intricate user queries, they often hallucinate non-existent product features or contradict their own earlier statements. These qualitative failures highlight the urgent need for enhanced fine-grained visual-text alignment and robust multi-step reasoning in e-commerce applications.

\section{Conclusion}

We presented \benchname{}, a multimodal benchmark for evaluating foundation models in real-world bilingual (Chinese--English) e-commerce, organizing 29 expert-verified tasks across 6 capability aspects for platform operators, merchants, and customers with visually salient and single-/multi-turn formats. Evaluating 20 representative models, we find that even frontier general-purpose models struggle, with the highest overall score below 70, while compressed rankings and failure modes expose shared weaknesses in fine-grained visual-text alignment, domain knowledge integration, and multi-step reasoning. \benchname{} thus serves as a diagnostic tool charting the path toward models that master the full e-commerce workflow rather than isolated recognition tasks.

\bibliographystyle{unsrt}
\bibliography{ref}


\newpage
\appendix
\section{Per-task Evaluation Results}
\label{appendix:per_task_results}

To complement the aggregated results in Section~\ref{sec:main_results}, Table~\ref{tab:per_task_results} reports the per-task scores of all evaluated models on the 29 tasks in \benchname{}. The final \emph{Average} row uses the same non-missing task-level macro average as the \emph{Overall} column in Table~\ref{tab:main_results}. This convention avoids treating unavailable multimodal results for text-only LLMs as zero, but it also means that models with many missing multimodal entries should be compared with that coverage difference in mind.

\subsection{Per-task Patterns}
\label{appendix:per_task_patterns}

The per-task view provides the evidence behind the compressed aggregate leaderboard in Table~\ref{tab:main_results}. Qwen3.5-397B-A17B appears in the top two on $11$ tasks, Gemini~3~Pro and Kimi~2.5 on $9$ tasks each, Gemini~3~Flash on $6$ tasks, and Qwen3-VL-32B on $5$ tasks. The leadership pattern is distributed rather than hierarchical: a model that is strong on review analysis or product understanding may still fail on personalization, multi-turn reasoning, or fine-grained image--text alignment. This highlights why an e-commerce benchmark must report task-level behavior rather than only an overall score.

Task difficulty varies substantially across the benchmark. Highlights Understanding has the highest task-average score ($99.0$), followed by E-commerce Knowledge Q\&A (Single-turn) ($96.3$), Product Category Classification ($89.3$), and UGC-to-Product Matching ($88.2$). These tasks mostly involve explicit extraction, recognition, or localized domain lookup. In contrast, Purchase Intention Prediction (Text) ($3.1$), Purchase Intention Prediction (Multi-modal) ($9.0$), MGC-to-Product Consistency Verification ($20.8$), and E-commerce Knowledge Q\&A (Multi-turn) ($23.3$) are the lowest-average tasks. The hard cases correspond to real commercial decisions where the model must synthesize user preferences, fine-grained visual evidence, and complex multi-step logic.

Notably, while models achieve high scores on single-turn tasks like EKSQA ($96.3$ average) that require isolated knowledge recall, they struggle severely on multi-turn tasks like EKMQA ($23.3$ average). Although these two tasks differ in construction and evaluation metrics and cannot be directly compared, their respective performance levels illustrate a broader trend: current models are proficient at retrieving static e-commerce facts but fail to maintain context, resolve coreferences, and synthesize information across dynamic, multi-turn interactions.

Across capability aspects, Kimi~2.5 is strongest on Product Understanding and Personalization Analysis, Qwen3-VL-32B leads Domain-Specific Knowledge and User Behavior Insight, and Gemini~3~Pro leads Review Analysis and the overall average. These aspect-level crossovers support the main-text claim that e-commerce capability is not reducible to a single general intelligence score. Compliance \& Quality Control requires more cautious interpretation because the newly added text-only LLMs only have the Address Verification score in this aspect; their high C\&QC averages therefore reflect the available text-only compliance result rather than full coverage of all six C\&QC tasks.

\section{Difficulty-wise VLM Performance}
\label{appendix:vlm_difficulty_performance}

\begin{figure}[t]
    \centering
    \includegraphics[width=0.98\textwidth]{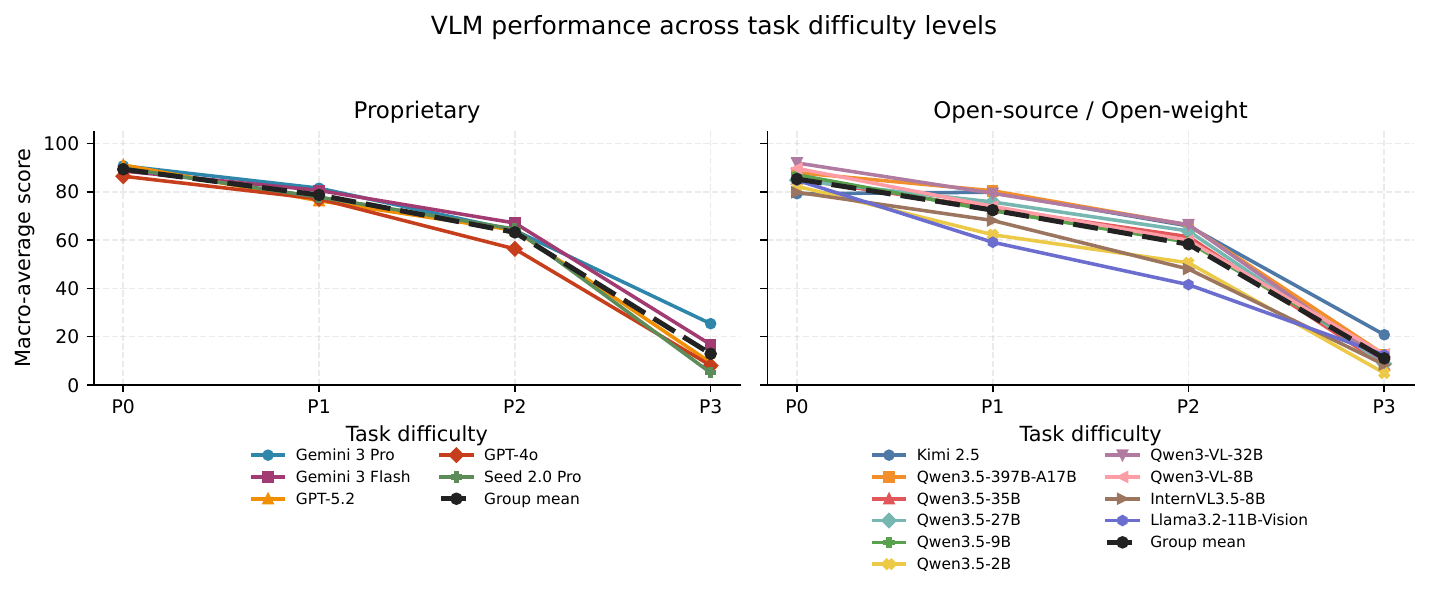}
    \caption{Difficulty-wise performance trends of all evaluated VLMs on \benchname{}. Each point is the non-missing macro average over tasks assigned to the corresponding P0--P3 difficulty tier. The dashed black curve reports the group mean within proprietary or open-weight VLMs.}
    \label{fig:vlm_difficulty_performance}
\end{figure}

Figure~\ref{fig:vlm_difficulty_performance} groups the per-task VLM results by the deterministic P0--P3 task labels introduced in Section~\ref{sec:design}. The aggregate trend is monotonic: the all-VLM macro average decreases from $86.6$ on P0 and $74.5$ on P1 to $59.9$ on P2, then drops sharply to $11.6$ on P3. This pattern indicates that the difficulty rubric effectively separates tasks that current VLMs can solve reliably from those that remain largely unsolved under zero-shot evaluation. The curve also explains the compressed overall leaderboard: models perform similarly well on fundamental extraction tasks, but complex e-commerce decisions expose shared bottlenecks.

\begin{sidewaystable}[t]
    \centering
    \caption{Per-task results on \benchname{} for all 29 tasks and the macro-average over available task scores (score in $[0,100]$, higher is better). Task abbreviations follow Table~\ref{tab:taxonomy}; model abbreviations follow Table~\ref{tab:main_results} (\emph{Gem}=Gemini, \emph{Seed-P}=Seed~2.0~Pro, \emph{DS-V4F}=DeepSeek-V4-Flash, \emph{Q3.5}=Qwen3.5, \emph{Q3VL}=Qwen3-VL, \emph{InVL}=InternVL, \emph{Lm}=Llama). \textbf{Bold} denotes the best score and \underline{underlined} the second-best within each row; -- denotes a missing result in the source spreadsheet.}
    \label{tab:per_task_results}
    \tiny
    \setlength{\tabcolsep}{2pt}
    \renewcommand{\arraystretch}{1.03}
    \resizebox{\textheight}{!}{%
    \begin{tabular}{llcccccccccccccccccccc}
    \toprule
    \textbf{Aspect} & \textbf{Task} & \textbf{Gem-3P} & \textbf{Gem-3F} & \textbf{GPT-5.2} & \textbf{GPT-4o} & \textbf{Seed-P} & \textbf{Kimi} & \textbf{Q3.5-397} & \textbf{Q3.5-35} & \textbf{Q3.5-27} & \textbf{Q3.5-9} & \textbf{Q3.5-2} & \textbf{Q3VL-32} & \textbf{Q3VL-8} & \textbf{InVL-8} & \textbf{Lm3.2V} & \textbf{GLM5.1-L} & \textbf{DS-V4F} & \textbf{GLM5.0-L} & \textbf{Lm3.3-70} & \textbf{Lm3.1-8} \\
    \midrule
    \multicolumn{22}{l}{\textit{Platform}} \\
    \cmidrule{1-22}
    \multirow{6}{*}{C\&QC} & PCV & 84.4 & \textbf{84.9} & 79.9 & 70.2 & 69.8 & 79.7 & \underline{84.5} & 77.4 & 81.9 & 71.9 & 59.8 & 74.4 & 79.9 & 59.3 & 54.8 & -- & -- & -- & -- & -- \\
     & ACFD & 46.5 & 76.0 & 71.5 & 66.0 & 71.5 & \underline{82.4} & \textbf{83.5} & 77.0 & 78.5 & 71.0 & 65.0 & 81.5 & 70.0 & 45.5 & 62.0 & -- & -- & -- & -- & -- \\
     & PDD & 92.5 & \textbf{98.3} & \underline{96.0} & 88.7 & 88.1 & 83.5 & 84.2 & 65.5 & 66.1 & 75.7 & 50.8 & 93.2 & 71.8 & 64.4 & 48.0 & -- & -- & -- & -- & -- \\
     & AV & \underline{97.9} & 94.0 & 92.9 & 86.9 & 93.9 & 96.6 & 93.0 & 89.0 & 85.0 & 77.0 & 46.0 & 95.0 & 81.0 & 78.8 & 59.0 & 92.0 & 92.3 & \textbf{99.0} & 88.0 & 58.0 \\
     & CCQA & \textbf{57.5} & 47.7 & 47.5 & 21.2 & \underline{55.6} & 31.4 & 33.9 & 35.0 & 47.7 & 45.6 & 24.4 & 53.8 & 48.8 & 27.5 & 24.4 & -- & -- & -- & -- & -- \\
     & MPCV & \textbf{67.9} & 45.0 & 12.0 & 8.5 & 9.5 & \underline{50.0} & 27.0 & 12.0 & 15.0 & 22.5 & 1.0 & 15.0 & 22.5 & 8.0 & 25.5 & -- & -- & -- & -- & -- \\
    \cmidrule(lr){1-22}
    \multirow{5}{*}{RA} & RSA & \textbf{87.6} & 82.3 & 82.3 & 83.0 & 82.5 & 83.8 & 82.4 & 77.6 & 78.0 & 76.2 & 74.4 & 85.3 & 76.7 & 76.9 & 75.1 & 75.5 & 79.5 & \underline{87.2} & 77.2 & 75.8 \\
     & RSR & \underline{62.0} & 61.9 & 60.7 & 58.8 & 57.0 & 58.7 & \textbf{62.6} & 56.3 & 61.4 & 56.7 & 51.6 & 60.2 & 57.9 & 44.9 & 40.9 & 61.2 & 58.5 & 60.5 & 56.3 & 41.2 \\
     & SLS & 85.5 & 82.0 & 74.5 & 83.0 & 79.0 & \underline{88.0} & \textbf{89.9} & 69.5 & 84.5 & 80.9 & 66.5 & 74.5 & 78.0 & 76.5 & 60.0 & 82.0 & 65.0 & 81.9 & 86.0 & 54.5 \\
     & MIRD & 83.8 & 79.6 & 84.8 & 77.2 & 81.5 & 66.1 & 78.5 & 72.5 & 72.9 & 77.4 & 70.0 & \underline{85.4} & 82.0 & 64.6 & 73.3 & 75.5 & 67.4 & 66.5 & \textbf{89.5} & 73.8 \\
     & RRG & \textbf{78.0} & \underline{75.0} & 49.0 & 45.0 & 55.0 & 48.0 & 54.1 & 54.5 & 56.0 & 42.0 & 27.0 & 42.0 & 49.0 & 37.0 & 34.0 & -- & -- & -- & -- & -- \\
    \midrule
    \multicolumn{22}{l}{\textit{Merchant}} \\
    \cmidrule{1-22}
    \multirow{6}{*}{PU} & SIM & 68.2 & 80.0 & 76.9 & 70.8 & 78.2 & 79.5 & \underline{83.0} & 80.7 & \textbf{83.7} & 75.2 & 66.0 & 80.3 & 76.1 & 62.7 & 50.8 & -- & -- & -- & -- & -- \\
     & PEC & 53.3 & 60.0 & 61.0 & 54.5 & \underline{66.7} & \textbf{79.8} & 64.4 & 42.9 & 50.3 & 49.5 & 42.2 & 62.9 & 54.6 & 32.4 & 18.9 & -- & -- & -- & -- & -- \\
     & PAQA & \textbf{42.4} & 37.9 & \underline{40.9} & 34.0 & 33.8 & 37.0 & 37.0 & 39.1 & 35.6 & 33.5 & 32.6 & 31.7 & 30.9 & 32.4 & 28.1 & -- & -- & -- & -- & -- \\
     & HU & 98.8 & \textbf{99.4} & 98.8 & 99.4 & 98.2 & \underline{99.4} & 99.4 & 98.8 & 99.4 & 99.4 & 99.4 & 98.8 & 99.4 & 98.2 & 98.8 & -- & -- & -- & -- & -- \\
     & UPRA & \textbf{88.7} & 75.3 & 47.3 & 56.1 & \underline{79.1} & 68.2 & 72.3 & 54.7 & 59.1 & 47.0 & 33.1 & 55.4 & 43.6 & 28.0 & 26.4 & -- & -- & -- & -- & -- \\
     & SPE & 84.9 & 84.9 & 83.2 & 81.3 & 80.8 & 85.1 & \underline{86.0} & \textbf{86.9} & 84.6 & 85.0 & 76.6 & 85.5 & 85.0 & 79.0 & 69.6 & -- & -- & -- & -- & -- \\
    \cmidrule(lr){1-22}
    \multirow{3}{*}{DK} & EKSQA & 97.7 & 97.6 & 97.4 & 95.7 & \underline{98.0} & 92.3 & 97.4 & 98.0 & 96.7 & 96.7 & 94.6 & \textbf{98.5} & 97.4 & 95.0 & 96.3 & 97.6 & 94.4 & 97.6 & 92.2 & 95.9 \\
     & EKMQA & 19.7 & 19.0 & \underline{31.8} & 18.1 & 19.4 & 20.6 & 24.1 & 22.3 & 24.0 & 22.2 & 18.9 & \textbf{68.4} & 22.9 & 16.5 & 14.9 & 22.1 & 26.2 & 18.6 & 20.0 & 15.4 \\
     & PCC & 91.8 & 91.9 & 92.0 & 88.4 & 89.3 & \textbf{93.7} & \underline{92.9} & 88.4 & 92.0 & 89.3 & 85.7 & 90.2 & 91.1 & 89.3 & 77.7 & -- & -- & -- & -- & -- \\
    \midrule
    \multicolumn{22}{l}{\textit{Customer}} \\
    \cmidrule{1-22}
    \multirow{4}{*}{UBI} & UVEC & 52.2 & 59.1 & 55.0 & 46.3 & 63.7 & \textbf{70.3} & \underline{65.2} & 56.2 & 54.5 & 52.6 & 48.3 & 61.3 & 53.3 & 35.3 & 19.4 & -- & -- & -- & -- & -- \\
     & UPM & 95.7 & \underline{96.9} & 95.5 & 92.6 & 95.8 & 95.1 & \textbf{97.2} & 88.5 & 92.7 & 87.1 & 79.8 & 93.4 & 87.1 & 80.5 & 45.6 & -- & -- & -- & -- & -- \\
     & QRQA & 69.7 & 76.4 & 71.5 & 70.7 & 74.8 & \underline{76.7} & \textbf{81.5} & 65.9 & 67.2 & 54.5 & 48.8 & 69.1 & 62.6 & 63.4 & 52.8 & 72.6 & 50.0 & 69.9 & 62.6 & 52.8 \\
     & CID & 53.5 & 55.0 & 55.0 & 65.0 & 52.5 & 55.5 & 52.0 & 64.0 & 67.5 & 63.5 & 69.0 & \underline{77.0} & \textbf{77.5} & 63.0 & 64.0 & 54.5 & 65.2 & 52.0 & 63.0 & 64.0 \\
    \cmidrule(lr){1-22}
    \multirow{5}{*}{PA} & TAA & 83.8 & 83.8 & 81.0 & 82.5 & 78.5 & 83.2 & 81.8 & 80.0 & 83.1 & \textbf{84.5} & 77.5 & \underline{84.5} & 84.5 & 82.5 & 68.0 & -- & -- & -- & -- & -- \\
     & PIPT & 1.7 & 2.3 & 2.8 & 3.4 & 0.6 & 1.5 & 3.8 & 4.0 & 3.4 & 2.8 & \textbf{5.1} & 3.4 & 4.0 & 3.4 & 4.5 & 1.1 & 3.8 & 1.7 & \underline{5.1} & 2.8 \\
     & PIPM & 6.5 & 2.9 & \underline{12.9} & 12.2 & 5.0 & 10.8 & 6.8 & 8.6 & 8.0 & 10.1 & 8.6 & 10.1 & 11.5 & \textbf{13.7} & 7.2 & -- & -- & -- & -- & -- \\
     & CPR & 80.1 & 80.9 & 79.1 & 80.4 & 78.4 & \underline{84.9} & \textbf{86.5} & 80.0 & 80.8 & 75.5 & 66.5 & 82.6 & 78.1 & 70.6 & 67.7 & -- & -- & -- & -- & -- \\
     & PRP & \textbf{71.1} & \underline{68.9} & 60.6 & 57.8 & 66.8 & 66.8 & 68.4 & 64.2 & 60.9 & 59.4 & 43.5 & 48.2 & 46.6 & 48.7 & 46.6 & -- & -- & -- & -- & -- \\
    \midrule
    \multicolumn{2}{l}{\textbf{Average}} & \textbf{69.1} & \underline{68.9} & 65.3 & 62.0 & 65.6 & 67.9 & 68.0 & 62.4 & 64.5 & 61.5 & 52.9 & 67.6 & 62.9 & 54.4 & 48.8 & 63.4 & 60.2 & 63.5 & 64.0 & 53.4 \\
    \bottomrule
    \end{tabular}%
    }
\end{sidewaystable}
\clearpage

The most severe degradation occurs at P3. Even the strongest P3 model, Gemini~3~Pro, reaches only $25.4$, while most VLMs remain near or below the low teens. The P3 tier contains tasks such as MGC-to-Product Consistency Verification and purchase-intention prediction, where success requires the joint synthesis of visual evidence, domain-specific rules, and user- or merchant-specific context. By contrast, Qwen3-VL-32B leads P0 with $92.0$, Gemini~3~Pro leads P1 with $81.5$, and Gemini~3~Flash leads P2 with $67.1$. This gap suggests that present VLMs handle explicit extraction and routine reasoning robustly, but fail at expert-level, multi-step decision-making.

The proprietary and open-weight curves exhibit similar trajectories, with no group immune to the P3 collapse. While open-weight VLMs are competitive on P0--P2, their performance also plummets at P3. This reinforces the main-text finding that aggregate leaderboard differences are overshadowed by the sheer difficulty of advanced tasks. Progress on \benchname{} will require models to acquire deep e-commerce grounding and robust multi-step reasoning abilities, rather than merely scaling up generic multimodal recognition.

\section{Task Examples by Difficulty Level}
\label{appendix:task_examples}

\begin{figure}[t]
    \centering
    \includegraphics[width=\textwidth]{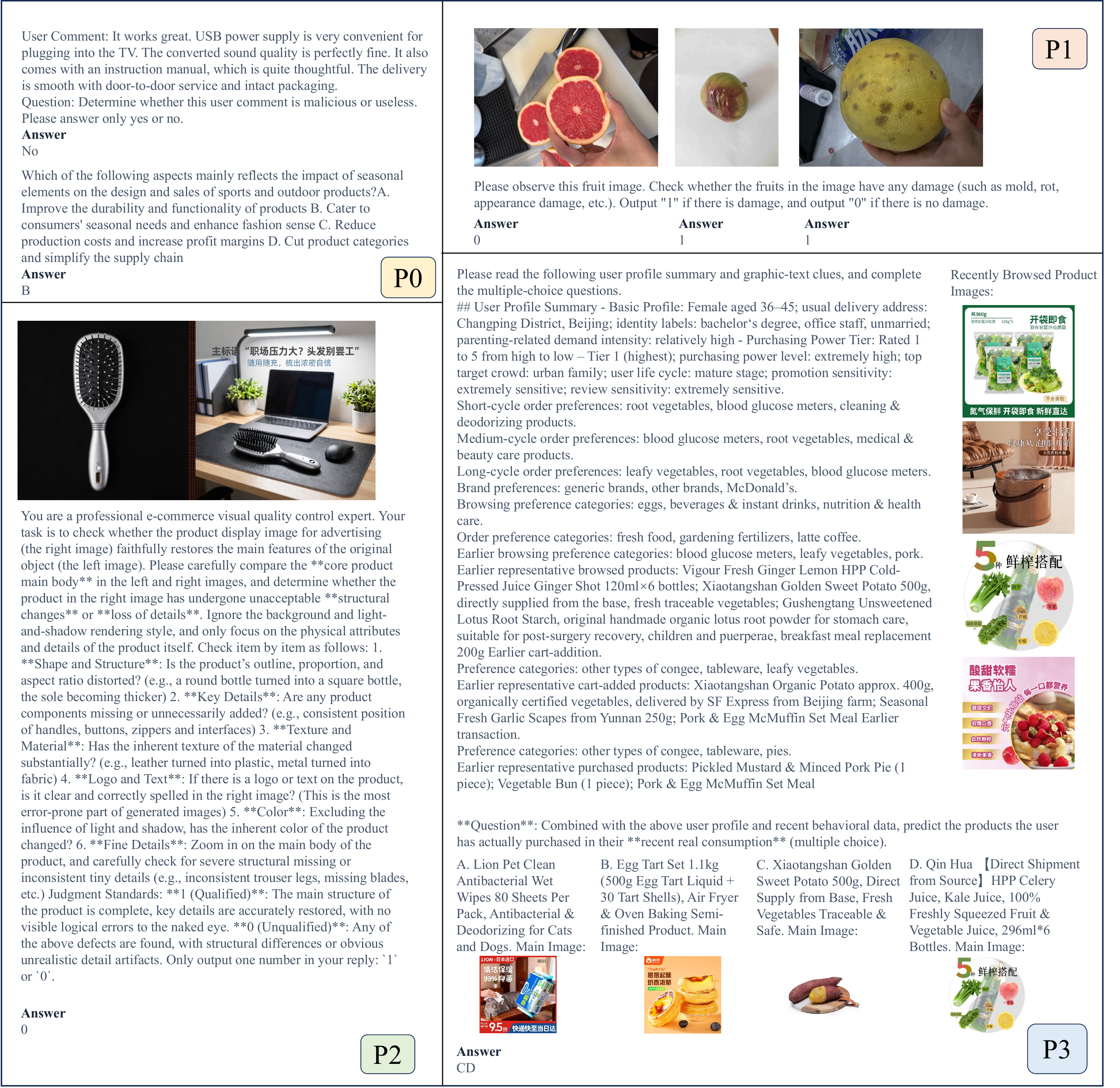}
    \caption{Representative task examples in \benchname{} across different difficulty levels. \textbf{Top-left (P0):} \emph{Malicious \& Invalid Review Detection} requires fundamental understanding to explicitly identify spam or invalid content in user reviews. \textbf{Top-right (P1):} \emph{Product Damage Detection} demands intermediate analysis and routine reasoning to identify specific defects or damage in product images. \textbf{Bottom-left (P2):} \emph{Ad Creative Fidelity Diagnosis} necessitates advanced reasoning and fine-grained visual inspection to detect unacceptable structural changes or detail losses between the original product and the generated advertisement image. \textbf{Bottom-right (P3):} \emph{Purchase Intention Prediction (Multi-modal)} involves expert-level judgment and comprehensive decision-making, requiring the model to synthesize detailed user profiles and long-term behavioral history to predict actual purchasing choices.}
    \label{fig:task_showcase}
\end{figure}

To complement the difficulty-aware categorization presented in Section~\ref{sec:design}, we provide concrete examples of representative tasks in \benchname{} across different difficulty levels (Figure~\ref{fig:task_showcase}). The four examples jointly cover the P0 to P3 difficulty tiers, illustrating the progression from fundamental understanding to comprehensive decision-making.

\section{Error Analysis of SOTA Multimodal Models}
\label{appendix:error_analysis}

To further understand the gap between open-domain capabilities and specialized e-commerce requirements, we conduct an error analysis on the failure cases of state-of-the-art (SOTA) multimodal large language models, taking Gemini~3~Pro as a representative example. As illustrated in Figure~\ref{fig:bad_cases}, we identify three primary challenges that general-purpose MLLMs currently face in e-commerce scenarios:

\begin{figure}[h]
    \centering
    \includegraphics[width=\textwidth]{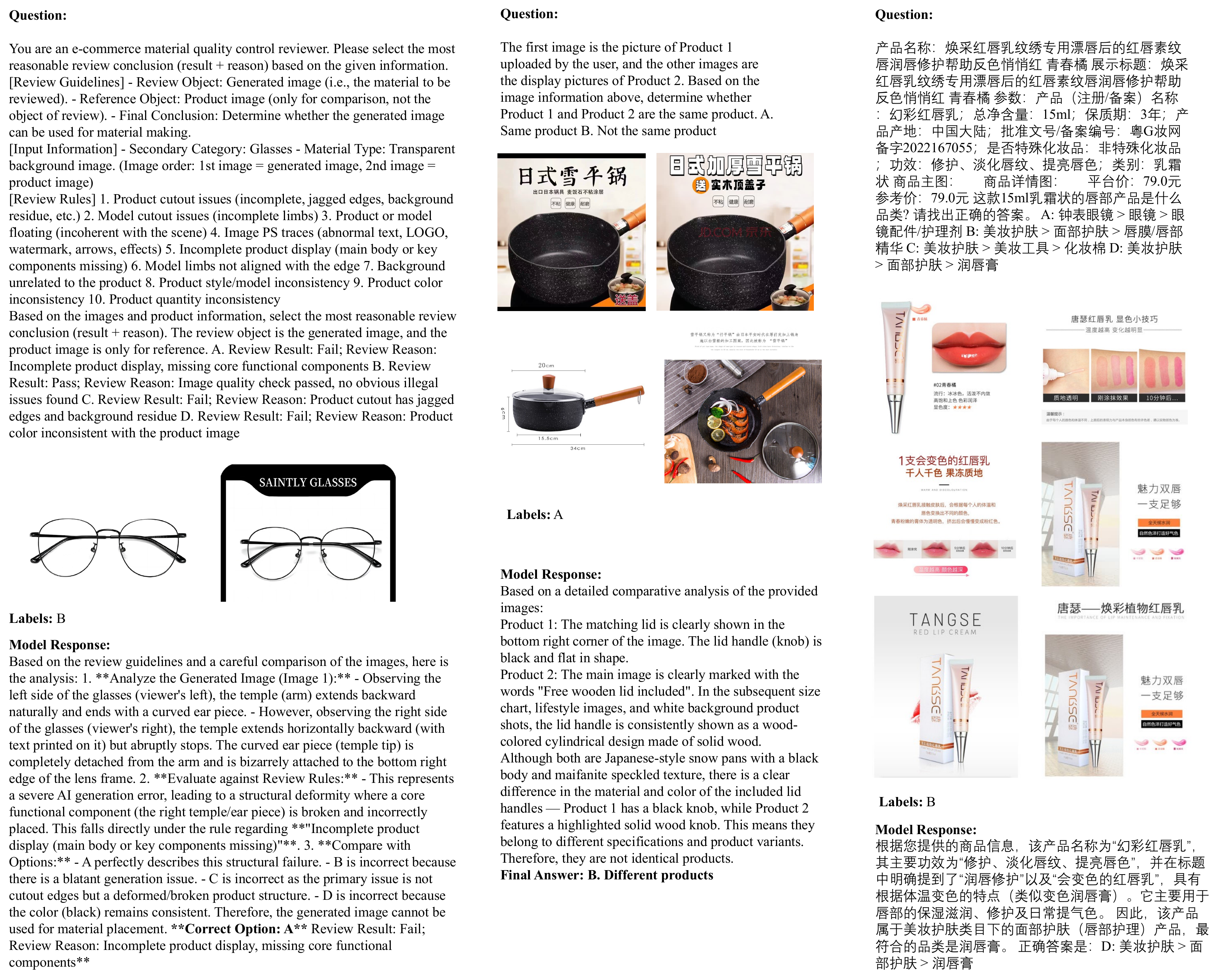}
    \caption{Representative failure cases of SOTA multimodal models (e.g., Gemini 3 Pro) in \benchname{}. The errors highlight three main deficiencies: severe visual hallucination in fine-grained recognition, confusion of domain-specific semantic concepts, and struggles with the dense and complex visual distribution typical of e-commerce images.}
    \label{fig:bad_cases}
\end{figure}

\textbf{1. Visual Hallucination and Fine-Grained Recognition Difficulties.} E-commerce applications necessitate highly precise visual inspection. In tasks like Product Damage Detection or MGC-to-Product Consistency Verification, models must detect subtle defects, texture differences, or minor missing accessories. General MLLMs frequently suffer from visual hallucinations, overlooking these fine-grained details and defaulting to a "normal" or "consistent" judgment, which is fatal for e-commerce quality control.

\textbf{2. Lack of Domain-Specific Professional Knowledge.} The e-commerce ecosystem operates on specialized terminology and distinct operational concepts. General MLLMs lack sufficient exposure to this deep professional knowledge. Consequently, in tasks like Target Audience Analysis or Selling Point Extraction, models frequently confuse similar semantic concepts or generate generic marketing copy that fails to capture the specific commercial intent or target demographic of the product.

\textbf{3. Complex Visual Distributions in E-commerce Scenarios.} Unlike standard natural images, e-commerce visuals (e.g., product detail pages, promotional posters, and user-generated reviews) are highly complex. They typically feature intricate layouts, dense text overlays, multiple sub-images, and varied professional photography styles. This significant distribution shift from natural images introduces substantial difficulties for general MLLMs, often causing them to miss key textual information embedded within promotional posters or fail to align multiple sub-images correctly.



\end{document}